# Standard and derived Planck quantities: selected analysis and derivations

Jason R. Buczyna

Department of Physics

University of Virginia

Charlottesville, VA 22904-4714, USA

jbuczyna@k12albemarle.org

C. S. Unnikrishnan

Gravitation Group, Tata Institute of Fundamental Research

Homi Bhabha Road

Mumbai – 400 005, India

unni@tifr.res.in

George T. Gillies

School of Engineering and Applied Science

University of Virginia

Charlottesville, VA 22904-4746, USA

gtg@virginia.edu

**Abstract** We provide an overview of the fundamental units of physical quantities determined naturally by the values of fundamental constants of nature. We discuss a comparison between the 'Planck units', now widely used in theoretical physics and the pre-quantum 'Stoney units' in which, instead of the Planck constant, the charge of the electron is used with very similar quantitative results. We discuss some of the physical motivation for these special units, attributed much after they were introduced, and also put forth a summary of the arguments supporting various cases for making specific physical interpretations of the meanings of some of these units. The new aspects we discuss are a possible physical basis for the Stoney units, their link to the Planck units, and also the importance of Planck units for thermodynamical quantities in the context of quantum gravity.

**Keywords** Planck units, Stoney units, early universe, quantum gravity, black hole thermodynamics, fundamental constants, SI units, dimensional analysis

**PACS numbers** 01.65.+g, 04.90.+e, 06.20.fa, 06.20.Jr

## 1 Introduction and background

As scientific knowledge advances, so does the interest in improved methods for defining and realizing the base units of the SI system, such as the second, the kilogram, the kelvin, etc. Realization of all the base units in terms of atomic quantities is a central goal of metrology because it allows for time-invariant and artifact-independent means of making and relating various physical measurements. Furthermore, in addition to providing for better standards of measurement, improvements in this area also help to refine the physical foundation for interpreting measurements and deepening our insights into their fundamental nature. In parallel with the needs of metrology are the needs of physics, and it is common practice among many workers to employ "natural units" in which for convenience of analysis or calculation, some of the physical constants are arbitrarily set have values of "1" (eg., $\hbar$, $c$, or $G_N$, the Planck constant, the speed of light, or the Newtonian gravitational constant, respectively).

The search for sets of fundamental units that can help us to better understand natural phenomena is not a new one. By *fundamental units* here and in what follows, we refer not to the base units of the SI system (even though by definition they are indeed the only universally accepted fundamental units of measurement extant today), but instead to some other type of special physical quantities that either can or do play an important role in one or more branches of physics. Physicists often point to the units developed by Planck[1] in 1899 as the first modern attempt at arriving at a system of fundamental units of this kind. However, such had already been accomplished[2] by Stoney in the 1870s and summarized[3] in an article in 1881, where he postulated a unified system of quantities that were dimensionally consistent with three of the most fundamental physical constants known at the time: Newton's gravitational constant $G_N$, the speed of light $c$, and the "quantity of electricity [that] traverses the electrolyte for each chemical bond which is ruptured" which later became the charge of the electron. With the modern values $e = 4.8\times10^{-10}$ esu., $G_N = 6.67\times10^{-8}$ dyne-cm$^2$/g$^2$, and $c = 2.998\times10^{10}$ cm/s, these units are given by

$$M_S = \sqrt{e^2 / G_N} \approx 2\times10^{-6}\ g \tag{1}$$

$$L_S = \sqrt{G_N e^2 / c^4} \approx 10^{-34}\ cm \tag{2}$$

$$T_S = \sqrt{G_N e^2 / c^6} \approx 3\times10^{-45}\ s\ . \tag{3}$$

The values estimated by Stoney were an order of magnitude smaller due to the smaller value of the charge used by him. The developments regarding these units are described in more detail in Barrow 2002.[4]

With the introduction and evolution of systems of units incorporating also those of the electromagnetism, natural units advocated by Stoney were not treated as a system of units for common use, even in theoretical discussions. The system of dimensioned ratios of fundamental constants that most prominently survives to the present day is that proposed by Planck in 1899 in the same paper where he also first proposed the fundamental constant that now bears his name. The constant that he called *b*, now universally referred to as *h*, is a fundamental constant that describes quanta of radiation and has units of angular momentum. Very importantly, it so happens that this constant (or, more commonly, its "reduced" form, $\hbar \equiv \frac{h}{2\pi}$) is not only fundamental in quantum mechanics but also in attempts to understand the nature of space-time itself and the dynamics of the early Universe. Planck used this constant in place of the electron charge, along with the speed of light and Newton's gravitational constant, to dimensionally derive fundamental units for the quantities of mass, length, and time. The modern understanding of these values (with *h* replaced by $\hbar$ as the fundamental unit of angular momentum) are as follows:

$$l_p = \sqrt{\frac{\hbar G_N}{c^3}} \approx 1.6 \times 10^{-35}\ m \tag{4}$$

$$M_{Pl} = \sqrt{\frac{\hbar c}{G_N}} \approx 2.2 \times 10^{-8}\ kg \tag{5}$$

$$t_p = \sqrt{\frac{\hbar G_N}{c^5}} \approx 5.4 \times 10^{-44}\ s \tag{6}$$

These are the three basic Planck units; they serve as the basis not only for many other possible Planck quantities, but more importantly, they set the scale for quantum gravity phenomena. As we will discuss here, there are many analogs to these units for other physical quantities, as well. We will also focus on the general potential of the Planck units to be used as a self-consistent set of fundamental and universal quantities for the study of physical phenomena, and perhaps as indicators of what may be an underlying set of even more profound physical quantities.

The organization of this paper will be as follows. First, we will make some general remarks about the possible relation between the Planck units, Stoney units and fundamental physics, with the well known and widely accepted speculations about the relation between Planck units and relativistic quantum gravity as the basis. Then we will analyze the known Planck units and their derivatives and, for some of these units, give examples of alternative derivations based on arguments more physically direct than dimensional analysis. Such arguments often support the belief that there is a deeper physical meaning than only one of universal convenience to many of the Planck units, and we will then discuss such interpretations in more detail as they arise.

Next, we will discuss the potential role that some of these units may play in arriving at an intrinsic physical system of units and, in conclusion, we will examine how sufficiently each of these points was addressed and emphasize the need for future work in this area.

## 2 Some remarks on the Planck units, Stoney units and fundamental physics

It is widely accepted and remarked, even without a fundamental theory, that the Planck quantities signify natural quantities associated with relativistic quantum gravity. In fact, all modern attempts to formulate a theory of quantum gravity takes the Planck length as a fundamental length scale associated with quantum gravity and goes to the extent of attributing this length scale to the fundamentals structure in the theory, like a string or loop length. Even space itself is supposed to be grainy at this length scale in the standard lore. Since the numerical values of the Stoney units are close to that of the Planck quantities, albeit without the signature constant $\hbar$ associated with quantum phenomena, physical interpretation of the Planck quantities as signifying quantum gravity might imply some deep connection between electromagnetism and relativistic gravity in the context of their quantum theories, perhaps signified in the relation $e^2 / c\varepsilon_0 = e^2 \left(\mu_0 / \varepsilon_0\right)^{1/2} \approx \hbar$. The relation between the two is important in the context of the possibility of cosmic variations of the fine structure constant, and also in the context of the quantum theory of spacetime, which is supposed to be quantitatively micro-structured in Planck units of length and time as a (3+1) dimensional 'foam'[5]. The fact, that the relation $e^2 \left(\mu_0 / \varepsilon_0\right)^{1/2} \approx \hbar$ is not exact also should not be glossed over, and the possible connection is at present a speculation. (Another instance when such a connection was speculated was in the context of the balance between the Casimir stress of a spherical 'shell' electron and the electrostatic stress[6]). If we include the energy dependent running of the electric charge, as measured and extrapolated, the relation is obeyed within a factor of 2 at the unification scale of $10^{16}$ GeV. We may expect that the relation might be exact at the Planck scale, thereby reducing the number of fundamental constants by one. However, if the gravitational coupling itself has an energy dependent evolution[7], new complexity will have to be considered.

A more natural setting for relating the Planck and Stoney units is the Kaluza-Klein theory and its variations. Since a higher dimensional theory like the Kaluza-Klein theory in 5D, with one compactifed dimension gives an explicit relation between the electric charge and the gravitational constant, ascribing physics behind Planck units equally well relates Stoney natural units with fundamental physics. In the 5D K-K theory,

$$e = \frac{hc\sqrt{2\kappa}}{R_0} = \frac{hc}{R_0}\sqrt{16\pi G_N / c^4} \tag{7}$$

where $R_0$ is the radius of the compactified dimension and $\kappa$ is the Einstein gravitational constant. The Stoney length is proportional to electric charge whereas the K-K compactification scale is inversely related to the charge. The naturalness and a possible physical basis of the Stoney fundamental length $L_S$ is then contained in the observation

$$\frac{L_S}{R_0} = \frac{e^2}{hc\sqrt{16\pi}} = \frac{\alpha}{8\pi^{3/2}} \tag{8}$$

or in the relation $(L_S R_0)^{1/2} \approx l_P$.

It is interesting to note that the Stoney mass $M_S = \sqrt{e^2 / G_N} \sim 2 \times 10^{-6}\ g$ allows the description of the gravitational coupling constant as a fundamental electric charge to gravitational charge ratio, $e / M_S = \sqrt{G_N}$ . However, this is useful from a metrology point of view only if the Stoney mass can be identified with some fundamental mass, which is not the case today.

In a recent work, Boya *et al* have noted[8] that a new system of natural units with rational functions (without square roots) could be discussed in which the gravitational constant $G_N$ of the Planck units is replaced by another constant $G_2$ defined through a fictitious force equation $F = G_2 mM / r$. In their work $G_2$ is defined such that the fundamental mass is the Planck mass itself. The Stoney units come very close to this requirement in that they are more 'rational' than the Planck units, with the square root factor only in the gravitational constant. However, we note that it is actually $\sqrt{G_N}$ the gravitational coupling, and not $G_N$ itself (from the gravitational force law, $F = \left(\sqrt{G_N} m_1\right)\left(\sqrt{G_N} m_2\right)/r^2$ (this is why the Planck mass is usually termed as the inverse gravitational coupling).

## 3 Analysis and derivations of Planck quantities

### 3.1 The base units

The original derivation of the Planck "base" units of mass, length and time was via dimensional analysis. Attributing physical basis for the existence of such quantities was motivated by the post-quantum field theory expectation of a quantum gravity theory. Apart from defining fundamental scales of length, mass, and time due to fundamental quantum gravitational constraints on space-time and matter, such an encompassing theory can be expected to provide a basis for microscopic physical description of all of physics. If the Planck quantities are actually fundamental physical quantities, then simple scaling laws would thus allow all other physical quantities to be expressed dimensionlessly in terms of them. This allows a discussion on the physics behind these quantities, including fundamental interrelationships. This is reminiscent of the relationships within the Dirac Large Numbers Hypothesis[9] (LNH). In part, the LNH states that large dimensionless quantities like the Eddington numbers at the heart of the LNH should be related to each other through simple algebraic relations with exponents of order unity. Then cosmological quantities got related to microscopic physics, and an evolving cosmology was the basis for discussing temporal variation of fundamental constants. Similarly, Planck quantities can be 'derived' and linked through an underlying physical argument, speculative as it may be, that is plausible and justifiable within a future encompassing theory.

A physical scaling argument for obtaining the Planck mass, other than through dimensional analysis arguments, was presented by de Sabbata[10] in 1994. The argument involves analyzing a superstring's tension, which is essentially the energy (or mass) per unit length, and noting that the maximum mass will occur for a string with length equal to the Planck length and at the point where symmetry breaking occurs. The tension in this regime is found to be

$$T_{Pl} = c^2 / G_N = T_{\max}\ , \ \left(M \le M_{Pl}\right). \tag{7}$$

Then, multiplying this by a length equal to the Planck length, which is thought to be the distance scale at which space-time becomes discontinuous, yields the Planck mass:

$$M_{Pl} = \sqrt{\hbar c / G_N}\ , \tag{8}$$

with the speed of light being to the fifth power under the radical if one is interested in the Planck energy.

In addition to Planck mass and Planck length, we also have the Planck time, which some argue is the smallest meaningful interval of time. Here too the value is typically arrived at via dimensional analysis. However, it may also be synthesized from quantum field theory arguments and other approaches. For instance, it has also been pointed out[11] that the Hawking formula for black hole evaporation comes about naturally due to the existence of a minimum meaningful unit of time. Others have generated it by adding torsion to general relativity: time can be defined on the quantum geometrical level through torsion as

$$t = \left(1/c\right)\oint \vec{Q}\, dA = n\sqrt{\hbar G_N / c^5}\ , \tag{9}$$

where $n$ is the normal quantum number and $Q$ is the torsion tensor (and $dA$ is thus a two-form). Hence, $n = 1$ corresponds to the minimum possible unit of time and happens to be equal to the Planck time. It follows that there is a limiting 'Planck frequency' defined by

$$f_{max} \approx (c^5/\hbar G_N)^{1/2}, \tag{10}$$

which is also important in realizing an *ad hoc* factor in renormalization in quantum field theories; with a value on the order of $10^{48}$ Hz, it is indeed very large, as required in order for the calculations to converge. See Horowitz and Wald for details of these arguments. One could potentially interpret this quantity as the highest physically possible rate of information flow for which bits may be discerned by a measuring device.

### 3.2 Other representative derived quantities

In a very clear and concise paper, Shalyt-Margolin and Tregubovich[12] derive the fundamental length, time, and inverse temperature, all while generalizing the uncertainty principle for energy and inverse temperature to extremely high-energy regimes. They begin with the generalized form of the position-time uncertainty principle, where Veneziano's limit[13] is employed:

$$\Delta x \geq \frac{\hbar}{\Delta p} + const\, l_p^2 \left( \frac{\Delta p}{\hbar} \right). \tag{11}$$

They then point out that Adler and Santiago showed[14] in 1999 that the constant term is equal to unity, allowing the equation to be cast in a simple quadratic form:

$$l_p^2 (\Delta p)^2 - \hbar \Delta x \Delta p + \hbar^2 \leq 0. \tag{12}$$

After noting that setting the discriminant in the previous equation equal to zero

$$(\hbar \Delta x)^2 - 4 \cdot l_p^2 \cdot \hbar^2 = 0 \tag{13}$$

to obtain the minimum value of the position uncertainty such that it still makes sense to talk about a momentum uncertainty, one finds that $\Delta x_{\min} = 2l_p$. Furthermore, one may also divide the starting equation by the speed of light $c$ and employ the same techniques to find that the minimum sensible time is $\Delta t_{\min} = 2t_p$. The authors also define the Planck momentum

$$p_{pl} = E_p / c = \sqrt{\hbar c^3 / G_N}. \tag{14}$$

They then consider the uncertainty principle between energy and inverse temperature and note that it must be generalized at high energies into the form

$$\Delta\left(\frac{1}{T}\right) \geq \frac{k}{\Delta U} + \eta\,\Delta U\,, \tag{15}$$

where for "dimension and symmetry reasons," $\eta = \frac{k}{E_p^2}$. The same analysis as before leads to the conclusion that

$$\beta_{\min} = \frac{1}{kT_{\max}} = \frac{\Delta t_{\min}}{\hbar}, \tag{16}$$

which can be inverted to yield the maximum sensible temperature:

$$T_{\max} = \frac{\hbar}{2t_p k} = \frac{\hbar}{\Delta t_{\min} k}. \tag{17}$$

They give an interpretation of the Planck temperature as the regime in which the deformed density matrix $\rho(\tau)$ (where $\tau$ is a parameter describing the deformation effects caused by curved space-time) must be considered to accurately describe thermodynamics in general relativistic space-time instead of the normal density matrix $\rho = \sum_n \omega_n |\phi_n\rangle\langle\phi_n|$. The Planck temperature is often interpreted as the maximum possible sensible temperature. This is also the point at which the gravitational energy of photons becomes significant.

The relation between Planck-scale gravity and Planck-scale thermodynamics has special significance in the context of quantum gravity theories and also some recent work on the possible emergence of gravitational physics from an underlying statistical physics and thermodynamics associated with the space-time[15]. This is the motivation for our treating temperature as a fundamental quantity, and not just energy scaled by the Boltzmann constant. Indeed, even in classical physics temperature signified both average energy as well as fluctuations, which assume special significance in the link between gravity and statistical mechanics.

A minimum sensible temperature is suggested by de Sabbata as arising from the existence of a minimum sensible acceleration, which also appears in MOND theories[16] and variations[17] thereof. He hypothesizes that the minimum sensible acceleration is given by

$$a_{\min} = cH_0 , \tag{18}$$

where $H_0$ is the present value of the Hubble constant, and he notes that the minimum sensible temperature is implied by this equation due to the relation

$$k_B T_{\min} c / \hbar = cH_0 \tag{19}$$

which agrees with the minimum value suggested by the time-temperature analog of the uncertainty principle[11,18]. However, we note that smaller accelerations are routinely recorded in gravitational measurements employing torsion balances and therefore, the interpretation of the acceleration $cH_0$ needs revision.

Continuing the development of these quantities, we note that one can add and analyze new Planck thermodynamic units by appealing to the quantities of 'Planck pressure', 'Planck volume', and the 'Planck temperature.' The Planck pressure can be interpreted as the maximum possible amount of pressure that the space-time can sustain. In that case, the 'Planck number of moles' is equal to just the inverse of the Avogadro number. To show this, we begin with the ideal gas law,

$$PV = nRT \tag{20}$$

giving $n = \frac{PV}{RT}$. One can then substitute in the expressions for Planck pressure, the Planck volume (*viz.*, the cube of the Planck length), and the Planck temperature to eventually arrive at the Planck mole. We start by noting that the "Planck pressure" would be equal to the Planck momentum transport in Planck time, across an area $l_P^2$.

$$P_{Planck} = \frac{M_{Pl} c / t_p}{l_p^2} = \frac{M_{Pl}}{t_p^2 l_p} \quad (21)$$

Then, since the Planck temperature is just the Planck energy divided by Boltzmann's constant,

$$T_{Pl} = \frac{M_{Pl} c^2}{k_B}, \quad (22)$$

we have the following for the "Planck number of moles":

$$n_{Pl} = \frac{\left(\frac{M_{Pl}}{t_p^2 l_p}\right)\left(l_p^3\right)}{R \cdot \frac{M_{Pl} c^2}{k_B}} = \frac{k_B \cdot l_p^2}{R \cdot t_p^2 c^2} = \frac{k_B}{R} = 1.6605388 \times 10^{-24} = 1 / N_A . \quad (23)$$

The important notion of Planck entropy is discussed in the next section. The Planck units are summarized in Table 1, along with representative references.

**Table 1**

Various Planck units and other examples of maximum and minimum sensible units.

| Planck unit | Formula | Value in SI units | Reference |
|---|---|---|---|
| ***Base Units:*** | | | |
| **Length** | $l_p = \sqrt{\frac{\hbar G_N}{c^3}}$ | $1.6162 \times 10^{-35}\ m$ | Planck |
| **Mass** | $M_{Pl} = \sqrt{\frac{\hbar c}{G_N}}$ | $2.1765 \times 10^{-8}\ kg$ | Planck |
| **Time** | $t_p = \sqrt{\frac{\hbar G_N}{c^5}}$ | $5.3912 \times 10^{-44}\ s$ | Planck |
| **Charge** | $q_{Pl} = \sqrt{4\pi\varepsilon_0 \hbar c}$ | $1.87554586 \times 10^{-18}\ C$ | Borzeszkowski and Treder [19] |
| *Fundamental Units:* | | | |
| Entropy | $k_B$ | $1.3806505 \times 10^{-23}\ J/K$ | fundamental |
| Velocity | c | $299792458\ m/s$ | fundamental |
| Action | $\hbar$ | $1.05457163 \times 10^{-34}\ J \cdot s$ | fundamental |
| *Energy*: | | | |
| $E_{\max}$ | $E_p = M_{Pl} c^2 = \sqrt{\frac{\hbar c^5}{G_N}}$ | $1.9561 \times 10^{9}\ J$ | Barrow |
| $E_{\min}$ | $\frac{\hbar c}{R_H}$ | $2.4 \times 10^{-52}\ J$ | de Sabbata |
| $m_{\min}$ | $E_{\min} / c^2$ | $2.7 \times 10^{-69}\ kg$ | (this paper) |
| *Temperature*: | | | |
| $T_{max}$ | $\frac{M_{Pl} c^2}{k_B} = \sqrt{\frac{\hbar c^5}{G_N k_B^2}}$ | $1.4168 \times 10^{32}\ K$ | Gorelik and Ozernoi[20] |
| $T_{min}$ | $\frac{\hbar c}{k_B R_H}$ | $1.8 \times 10^{-29}\ K$ | de Sabbata |
| *Acceleration*: | | | |
| $a_{max}$ | $\frac{c}{t_p} = \sqrt{\frac{c^7}{\hbar G_N}}$ | $5.5608 \times 10^{51}\ m/s^2$ | Falla and Landsberg[21] |
| $a_{min}$ | $k_B T_{\min} c/\hbar = cH_0$ | $6.9 \times 10^{-10}\, m/s^2$ | de Sabbata |

*Momentum:*

| | | | |
|---|---|---|---|
| *Planck* | $m_{Pl} c = \sqrt{\frac{\hbar c^3}{G}}$ | $6.5248\ kg \cdot \frac{m}{s}$ | Casher and Nussinov[22] |
| $p_{min}$ | $m_{\min} c$ | $8.1 \times 10^{-61}\ kg \cdot \frac{m}{s}$ | (this paper) |

*Other (Derived) Planck Units:*

| | | | |
|---|---|---|---|
| Frequency | $t_p^{-1} = \sqrt{\frac{c^5}{\hbar G_N}}$ | $1.8549 \times 10^{43}\ Hz$ | Horowitz and Wald[23] |
| Area | $l_p^2 = \frac{\hbar G_N}{c^3}$ | $2.6122 \times 10^{-70}\ m^2$ | Zhu[24] |
| Volume | $l_p^3 = \left(\frac{\hbar G_N}{c^3}\right)^{\frac{3}{2}}$ | $4.2220 \times 10^{-105}\ m^3$ | Hawking[25] |
| Density | $\frac{M_{Pl}}{l_p^3} = \frac{c^5}{\hbar G_N^2}$ | $5.1550 \times 10^{96}\ kg/m^3$ | Harrison[26] |
| Power | $M_{Pl} c^2 / t_p = \frac{c^5}{G_N}$ | $3.6283 \times 10^{52}\ W$ | Gerlach[27] |
| Force | $F_{Pl} = M_{Pl} \frac{c}{t_p} = \frac{c^4}{G_N}$ | $1.2103 \times 10^{44}\ N$ | Winterberg[28] |
| Pressure | $\frac{F_{Pl}}{l_p^2} = \frac{c^7}{\hbar G_N^2}$ | $4.6331 \times 10^{113}\ Pa$ | (this paper) |
| "Mole No." | $n_{Pl} = \frac{P_{Pl} V_{Pl}}{R T_{Pl}}$ | $1.6605388 \times 10^{-24}$ | (this paper) |
| Viscosity | $\frac{F_{Pl} t_p}{l_p^2} = \sqrt{\frac{c^9}{\hbar G_N^3}}$ | $1.6671 \times 10^{60}\ Pa \cdot s$ | Gibson[29] |
| Voltage | $V_{Pl} = E_{Pl} / q_{Pl} = \sqrt{\frac{c^4}{4\pi G \varepsilon_0}}$ | $1.0429 \times 10^{27}\ V$ | Lundgren[30] |
| Current | $I_{Pl} = q_{pl} / t_p = \sqrt{\frac{4\pi\varepsilon_0 c^6}{G_N}}$ | $3.4789 \times 10^{25}\ A$ | Lundgren |
| Resistance | $\Omega_{Pl} = \frac{V_{Pl}}{I_{Pl}} = \frac{1}{4\pi\varepsilon_0 c}$ | $29.9792458\ \Omega$ | Lundgren |
| Capacitance | $C_{Pl} = 4\pi\varepsilon_0 \sqrt{\frac{\hbar G_N}{c^3}}$ | $1.7983 \times 10^{-45}\ F$ | Lundgren |
| Inductance | $L_{Pl} = \frac{1}{4\pi\varepsilon_0} \sqrt{\frac{\hbar G_N}{c^7}}$ | $1.6162 \times 10^{-42}\ H$ | Lundgren |
| Electric field | $\sqrt{\frac{c^7}{4\pi\varepsilon_0 \hbar G_N^2}}$ | $6.4529 \times 10^{61}\ V/m$ | Lundgren |
| Magnetic field | $\sqrt{\frac{c^5}{4\pi\varepsilon_0 \hbar G_N^2}}$ | $2.1525 \times 10^{53}\ T$ | Magueijo 1993 |

## 4 Discussion

The quantities listed in Table 1 have often been considered in terms of their potential for establishing a natural and unambiguous system of units that derive from the fundamental structure and laws of the physical universe. The first three of the base units, the Planck length, Planck mass, and Planck time, would clearly serve as the base units in this alternate metrology because they are expressed only in terms of constants $\hbar$, $c$ and $G_N$ that are thought to be invariant throughout space and time. The Planck charge incorporates the permittivity of free space, and is therefore somewhat less fundamental than the first three because of this need for an additional constant, but nevertheless it is included in most assessments of the Planck base units. From that point, the remaining units are all "derived units," in analogy with the derived units of the SI, because they are generated through manipulation of the base units and the fundamental constants.

If one takes the position that the fundamental unit of temperature be given by the fundamental energy unit divided by the fundamental entropy unit, then this leads directly to the Planck temperature and this quantity can then be understood objectively: the triple-point of water will be known as $1.93\times10^{-30}$ of the fundamental temperature throughout the Universe. From this, the fundamental unit of entropy is still the Boltzmann constant (or, more precisely, the Boltzmann constant multiplied by ln 2), and its value would be the fundamental temperature divided by the fundamental energy. Thus, if one could derive the fundamental temperature by completely independent means, i.e., via a method not dependent upon the Boltzmann constant, then one would have a method for enumerating this constant. Nonetheless, the fact that temperature as it arises within this context may be considered to be a fundamental unit possibly implies that there may be a connection between the concept of temperature and space-time itself on the same level as the connection between mass, energy, and dimensions.

Indeed, this is an important point requiring further work. It is well known that many derivations inspired by quantum gravity use the method of converting Lorentzian space-time to Euclidean space-time, by Wick rotation, implying a mapping between time and temperature in certain situations. In the context of black hole quantum physics, the fundamental entropy, or the Boltzmann constant, can be interpreted as the quantum entropy associated with the black hole of radius equal to the Planck length, approximately, from the relation for the black hole area entropy, $S_{BH} = k_B A / 4 l_{pl}^2$. Thus *the Boltzmann constant is the Planck scale entropy* and it can be interpreted as the *limit of information loss* in an evaporating black hole, anticipating that when the black hole evaporation reduces the black hole to one of Planck size, the minimal length, evaporation stops and a quantum steady state emerges, just as the ground state of a harmonic oscillator with a minimum nonzero energy. Thus, the Boltzmann constant represents the *asymptotic lossless information per black hole – the quantum minimum of entropy or the zero-point entropy*.

Now that there is a method of using the fundamental entropy unit as a special unit, one might also accept the *minimum* temperature, $\frac{\hbar c}{k_B R_H}$, as a fundamental unit. Although this might not seem unreasonable, it does not seem desirable for our analysis. This is because the Hubble radius (which would need to be expressed in quantities of the Planck length) is constantly changing, so the only real use for this as a fundamental unit would be if one of the quantities *c*, $\hbar$, or $k_B$ also had a temporal dependence, and there is currently no evidence for this phenomenon. (Another possibility is that the minimum sensible temperature is related to the average of temperature of the universe, and then there will be a dependence on $R_H$ ). Thus, we will not further consider the minimum temperature unit in our analysis; similarly, we will also not further consider the minimum energy, mass, or momentum.

Thereafter, in Table 1, we then encounter a number of quantities that, even though they are based entirely on $G_N$, $c$, and $\hbar$, are nevertheless derived units. These are the Planck frequency, area, volume, density, power, force, and pressure. Some of these units are more easily understood than others because their derivations are simple, e.g., for the Planck density, one simply divides the base mass by the base volume. In the context of loop quantum gravity, the fundamental geometrical quantities like the area and volume also have special significance, being quantized[31]. This, when combined with black hole thermodynamics implies that entropy is also quantized, in units of $\eta k_B$ where $\eta \approx 1$. The "Planck number of moles" is one step removed from this scenario because it involves the value of the ideal gas constant, *R*.

Most of the rest of the entries in Table 1 are Planckian analogs of electromagnetic units and quantities involving eg., voltage, capacitance and field strengths. We note, however, that one could potentially derive many new quantities involving these concepts, for instance a current of one electron charge per Planck time. However, a current of 1 A would consist of $1.16 \times 10^{62}$ of these units, so it is hard to see what the motivation for derivation and use of such a quantity might be.

## 5 Summary and future potential

We have examined several of the more well known Planck quantities and have arrived at some that are little discussed if not unknown in the literature. We have summarized how others have arrived at alternative derivations for many of these, showing that it is possible to synthesize these quantities via methods other than dimensional analysis. The results were tabulated into an extensive listing of the Planck units and quantities, as per Table 1.

We have also examined the Planck quantities in terms of their potential use for a unified system of natural units and found that many, but not all, of the quantities that we have examined can be classified as either base units or derived units for that purpose. Of particular note is our discussion of Planck temperature as a fundamental unit, and the Boltzmann constant as the Planck entropy, motivated by past and recent work on the connections between statistical mechanics, thermodynamics, gravity, and the physics of space-time. We acknowledge that some of the Planck quantity analogs presented here do not have a direct physical motivation underlying their synthesis, e.g., the electromagnetic analogs fall especially in this category. However, the interesting exploratory nature of developments in this area, and the general motivation that a new metrological scale would enjoin, nevertheless make it useful to test the limits of the symmetries between the Planck units and the existing classes of anthropocentric ones.